\documentclass[a4paper]{jpconf}
\usepackage{graphicx}
\usepackage{amssymb,mathtools}
\usepackage{capt-of}
\begin{document}
\title{Influence of the spreading resistance on the conductance spectrum of planar hybrid thin film SNS' junctions based on iron pnictides}

\author{S~D\"{o}ring$^1$, S~Schmidt$^1$, S~Gottwals$^1$, F~Schmidl$^1$, V~Tympel$^1$, I~M\"{o}nch$^2$, F~Kurth$^2$, K~Iida$^2$, B~Holzapfel$^2$ and P~Seidel$^1$}

\address{$^1$ Friedrich-Schiller-University Jena, Helmholtzweg 5, 07743 Jena, Germany}
\address{$^2$ IFW Dresden, Helmholtzstr. 20, 01069 Dresden, Germany}

\ead{sebastian.doering.1@uni-jena.de}
\begin{abstract}
To investigate the superconducting properties of iron pnictides we prepared planar hybrid SNS' junctions in thin film technology with a pnictide base electrode, a gold barrier layer and a lead counter electrode. Our design allows characterization of the electrodes and the junction independently in a 4-probe method. We show how both electrodes influence the measured spectra due to their spreading resistance. While the Pb electrode has a constant resistance above its $T_c$, the contribution of the pnictide electrode is clearly current-dependent and thus it needs a more advanced method to be corrected. We present an empirical method, which is simple to apply and allows to deal with the spreading resistance in our junctions to recalculate the actual conductance and voltage of one junction at given temperature.
\end{abstract}
\section{Introduction}
Within Andreev reflection studies the conductance of a prepared junction or a point contact is measured in dependence on the voltage. Usually the results are modelled within the BTK-theory \cite{Blonder1982} and its possible extensions like scattering \cite{Dynes1978,Plecenik1994}, anisotropic order parameter \cite{Kashiwaya1996} or non-isotropic Fermi surfaces \cite{Mazin1999,Brinkman2002}. To apply the BTK-model one has to normalize each conductance spectrum at a given temperature by a spectrum measured at $T\gtrsim T_c$. A problem, which occurs especially when thin films are used is the spreading resistance of the electrodes. This leads to a downshift of the conductance for higher temperatures, which avoids simple normalization and results in an overestimation of gap values from the spectra \cite{Woods2004,Chen2010}. Additionally, the exceeding of the critical current of one electrode causes dips in the spectra \cite{Sheet2004,Baltz2009}. In our former work \cite{Doering2012b,Schmidt2012} we clearly obtained an influence of the electrode resistances. Thus, it is necessary to characterize both electrodes to subsequently identify the extent of influence on the junction spectra and correct the measured data within an appropriate model.

\section{Measurements setup and results}
\begin{figure}[hbtp]
\begin{minipage}[t]{0.42\textwidth}
\centering
\includegraphics[width=1\textwidth]{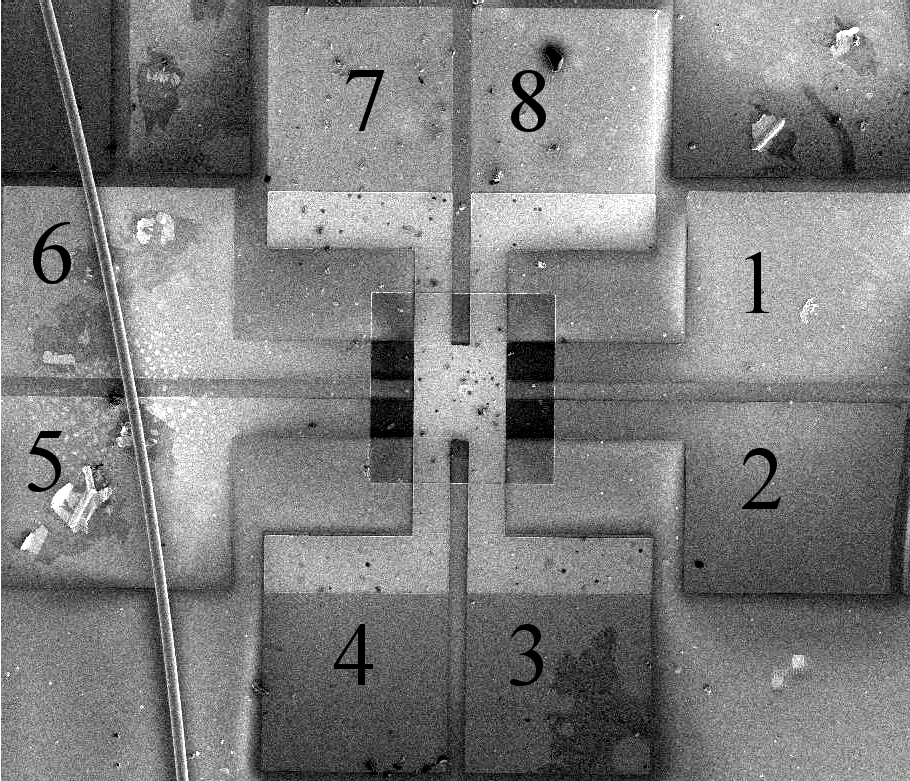}
\end{minipage}
\hspace{0.02\textwidth}
\begin{minipage}[b]{0.55\textwidth}
\centering
\caption{\label{fig:junction}Scanning electron microscope image of a planar SNS' cross type junction. The numbers mark the bonding pads. Their assignment to the electrodes are explained in table \ref{tab:contacts}. 
}
\vspace{.5cm}
\captionof{table}{\label{tab:contacts}Device parts corresponding to bonding pads numbers of figure \ref{fig:junction}.}
\begin{tabular}{l|l|l|l}
\textbf{setup} & \textbf{device part} & \textbf{current} & \textbf{voltage} \\
\hline
\textbf{I} & 122 electrode & 1-5 & 2-6 \\
\hline
\textbf{II} & Pb electrode & 8-4 & 7-3 \\
\hline
\textbf{III} & junction & 1-4 & 2-3 \\
\end{tabular}
\vspace{.1cm}
\end{minipage}
\end{figure}

\begin{figure}[hbtp]
\begin{minipage}[t]{0.47\textwidth}
\centering
\includegraphics[width=1\textwidth]{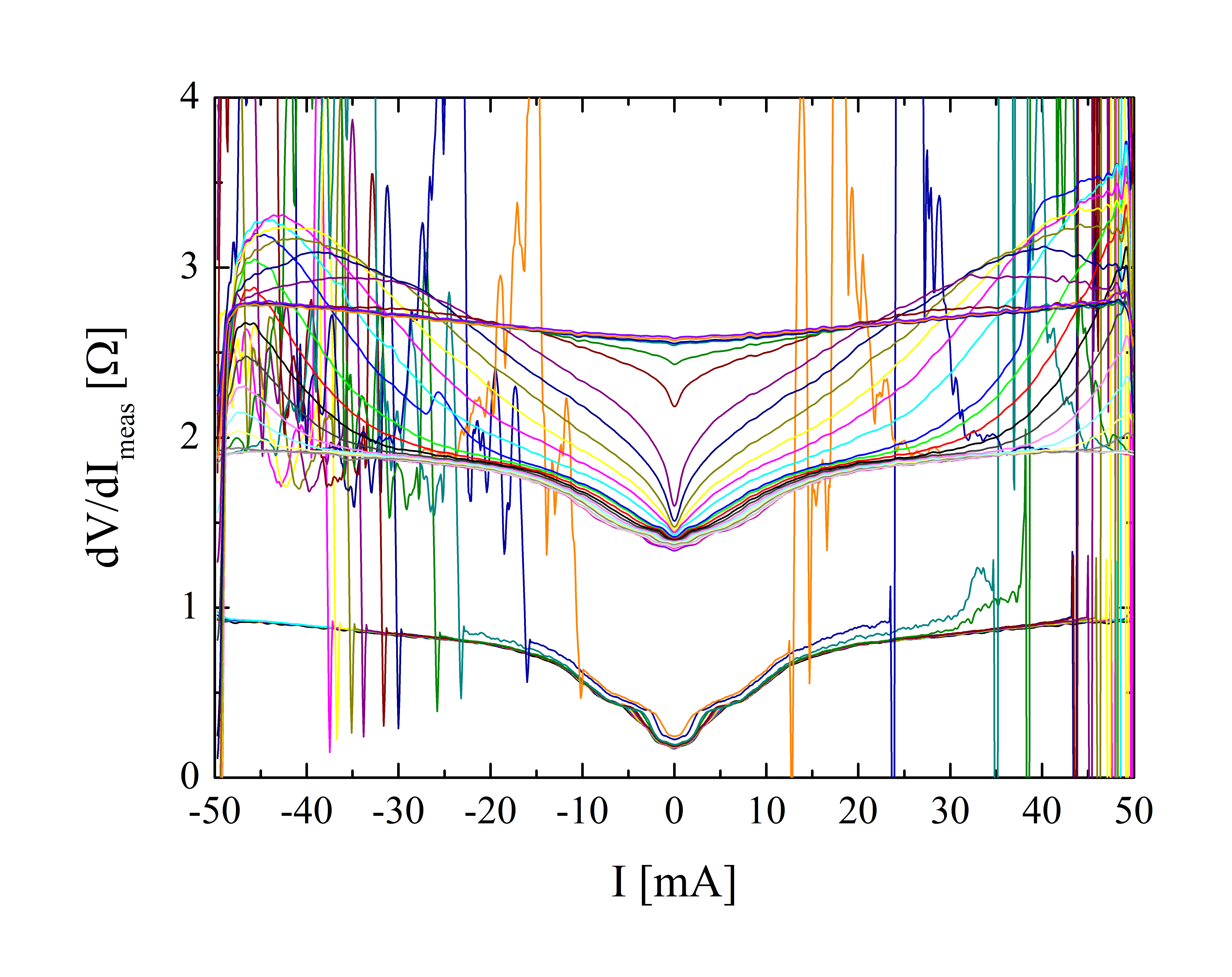}
\caption{\label{fig:R-I_uncorr}Differential resistance versus bias current of a junction at temperatures between 4.2\,K and 23.0\,K.}
\end{minipage}
\hspace{0.04\textwidth}
\begin{minipage}[t]{0.47\textwidth}
\centering
\includegraphics[width=1\textwidth]{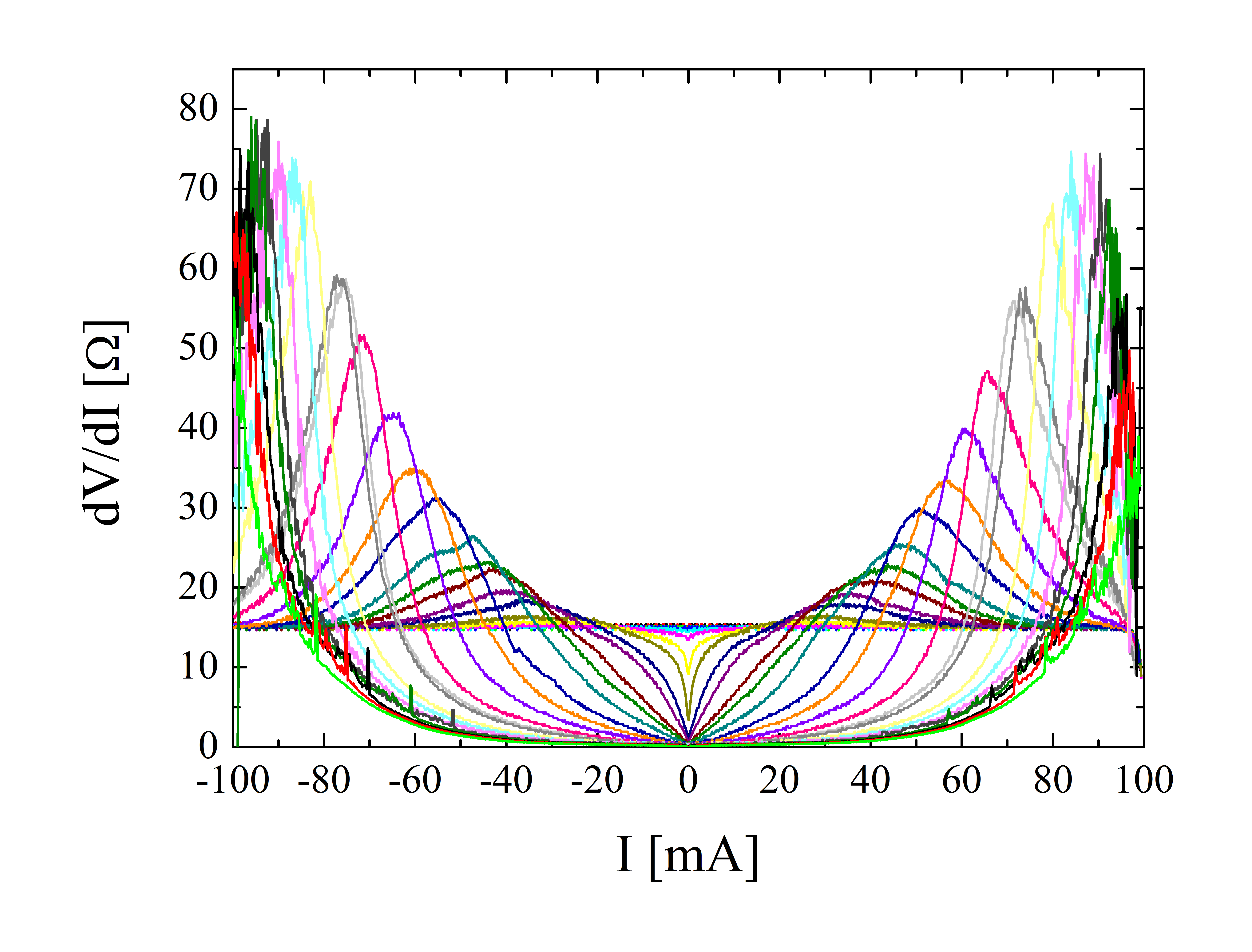}
\caption{\label{fig:R-I_122}Differential resistance versus bias current of the Ba-122 base electrode of the same junction at temperatures between 4.2\,K and 23.0\,K.}
\end{minipage}
\end{figure}

For our junctions we use a photolithographic mask design which allows to measure and characterize each electrode and the junction, respectively, in a 4-probe technique independently from each other. The used contacts for each of these measurements are shown in table~\ref{tab:contacts}. The details of junction preparation can be found in \cite{Doering2012b}.\\
\\
To measure the junction we used setup III from table~\ref{tab:contacts} to obtain the differential resistance over the bias current. The results are shown in figure~\ref{fig:R-I_uncorr}. One can distinguish three different regimes depending on the corresponding temperature. For temperatures lower than the $T_c$ of lead the spectrum shows a central dip with multiple shoulders for low currents. The resistance for high currents saturates at 1\,$\Omega$. For increasing temperatures jumps in the resistance occur, which correspond to the critical current of the counter electrode. At 7.5\,K the measured spectrum shows no jumps any more and it is shifted parallel by $\approx$1\,$\Omega$ upwards. The general shape stays nearly unchanged except for features caused by the gap of lead in the low temperature measurements. This additional 1\,$\Omega$ is due to the resistance of the lead counter electrode.\\
\\
By increasing the temperature further one can see an additional rise of the resistance for high currents. As can bee seen, the resistance in the normal state is about 2.8\,$\Omega$. This means, the Ba-122 base electrode causes another 0.8\,$\Omega$ resistance in the junctions spectrum. The rise in resistance increases, leading to peaks at special current values and for even higher currents a decrease of resistance, which closes asymptotically to the normal state value. Additionally, by increasing temperature the shape around zero current changes from a "u-shaped" to more "v-shaped" or mathematically from a convex function to a concave one.\\
By switching to setup II of table~\ref{tab:contacts} one can characterize the lead counter electrode. The differential resistance for $T\leq T_c$ shows zero resistance up to a critical current. At this critical current jumps occur at the same currents as the jumps in the junction's resistance when the measurement is done at the same temperature and the same bias direction. This is necessary because of the hysterical behaviour of the electrodes curve. For $I\geq I_c$ the differential resistance has a constant value, for a fixed $T\,>\,T_c$ it is current independent, too, but slightly increases with temperature.\\
On the same junction we can switch the electrical contacts to setup I of table~\ref{tab:contacts} to measure the behaviour of the Ba-122 base electrode at each temperature used for the junction's characterization. A typical behaviour visible in figure \ref{fig:R-I_122} is the occurrence of peaks in the differential resistance, which are mainly caused by flux flow and pinning but also possibly influenced by iron buffer layers \cite{Iida2010}. The height and the current position decrease with increasing temperature. One can identify these peaks in figure~\ref{fig:R-I_122} with those in figure~\ref{fig:R-I_uncorr}, because at given temperature the current position of one peak is the same in both figures. This fact together with the change from "u-shaped" to "v-shaped" current dependence of the differential resistance gives strong evidence of an influence of the Ba-122 electrode to the measured junction spectra. The resistance of the electrode increases from zero to 16\,$\Omega$ when increasing the temperature from 4.2\,K to 23\,K.  

\section{Model}
The correction algorithm we use is based on the one described in \cite{Baltz2009}. For the constant resistance of lead, this is just subtracted from the the measured junction resistance. In the same way the voltage of the junction is corrected by subtracting the Pb electrode's resistance times the bias current from the measured voltage. The value of this resistance can easily be obtained from figure \ref{fig:R-I_uncorr} as the first up shift of 1\,$\Omega$.\\
For the Ba-122 electrode in contrast to \cite{Baltz2009} we do not assume a constant spreading resistance but use its measured (current dependent) resistance characteristics. The extent of the influence $r_E\coloneqq {\Delta R_{junction}/\Delta R_{122}}$ can be obtained by comparing the raise of the normal resistance with temperature in figure \ref{fig:R-I_uncorr}  and figure \ref{fig:R-I_122}. As aforementioned, in the shown example $\Delta R_{junction}=0.8\,\Omega$ while $\Delta R_{122}=16\,\Omega$, thus giving $r_E=5\,\%$. Now the actual voltage and resistance of the junction can be calculated by subtracting the measured resistance dependence of the Ba-122 electrode times $r_E$ from the measured voltage and resistance, respectively for each temperature:
\begin{eqnarray}
V_{corr} & = & V_{meas} - I\cdot R_{Pb} - r_E\cdot R_{122}(I)\cdot I \label{eq:Vcorr}\\
\left(\frac{dV}{dI}\right)_{corr} & = & \left(\frac{dV}{dI}\right)_{meas} - R_{Pb} - r_{E} \cdot R_{122}(I) \label{eq:Rcorr}
\end{eqnarray}

\section{Corrected data}
\begin{figure}[hbtp]
\begin{minipage}[t]{0.47\textwidth}
\includegraphics[width=1\textwidth]{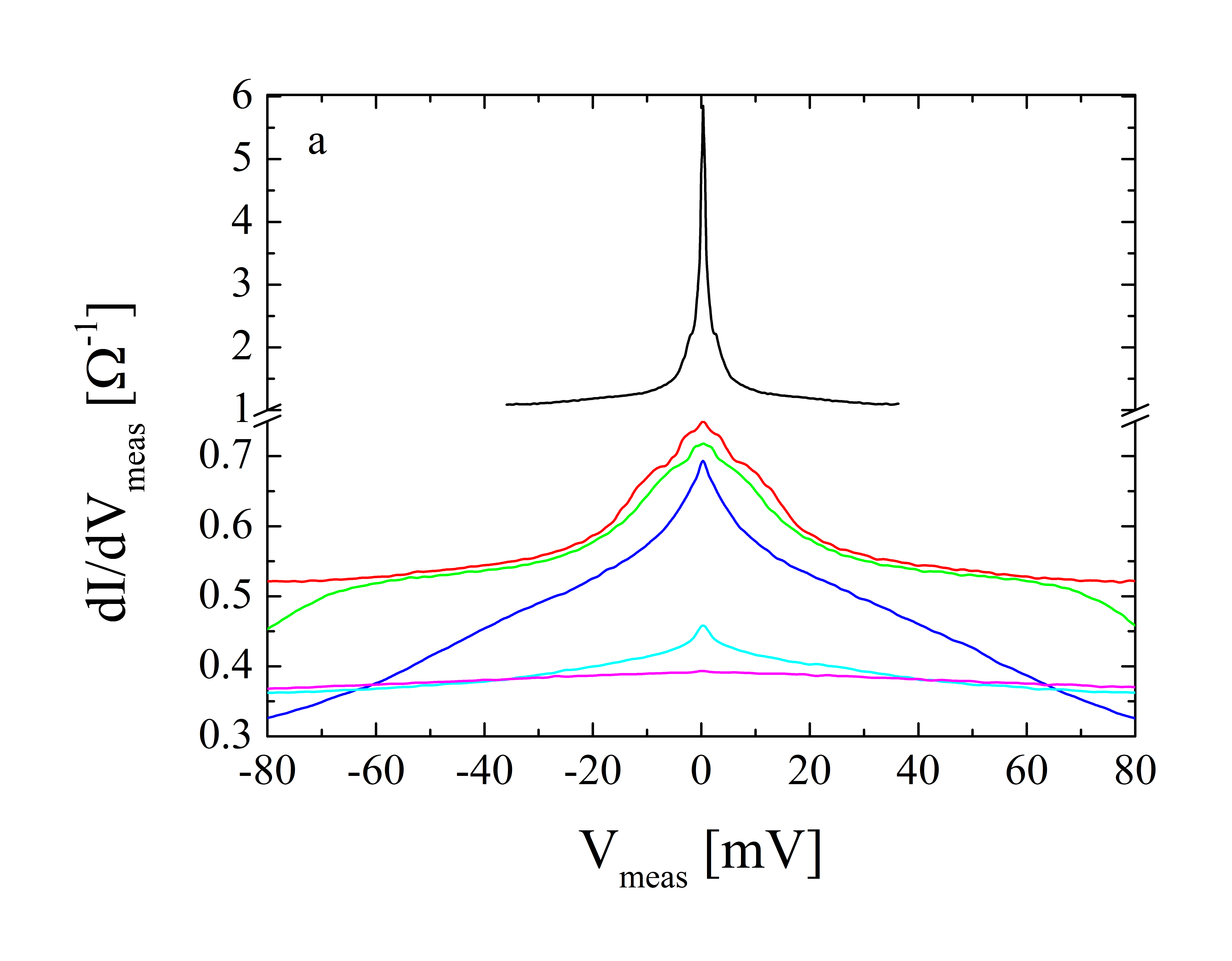}
\end{minipage}
\hspace{0.04\textwidth}
\begin{minipage}[t]{0.47\textwidth}
\includegraphics[width=1\textwidth]{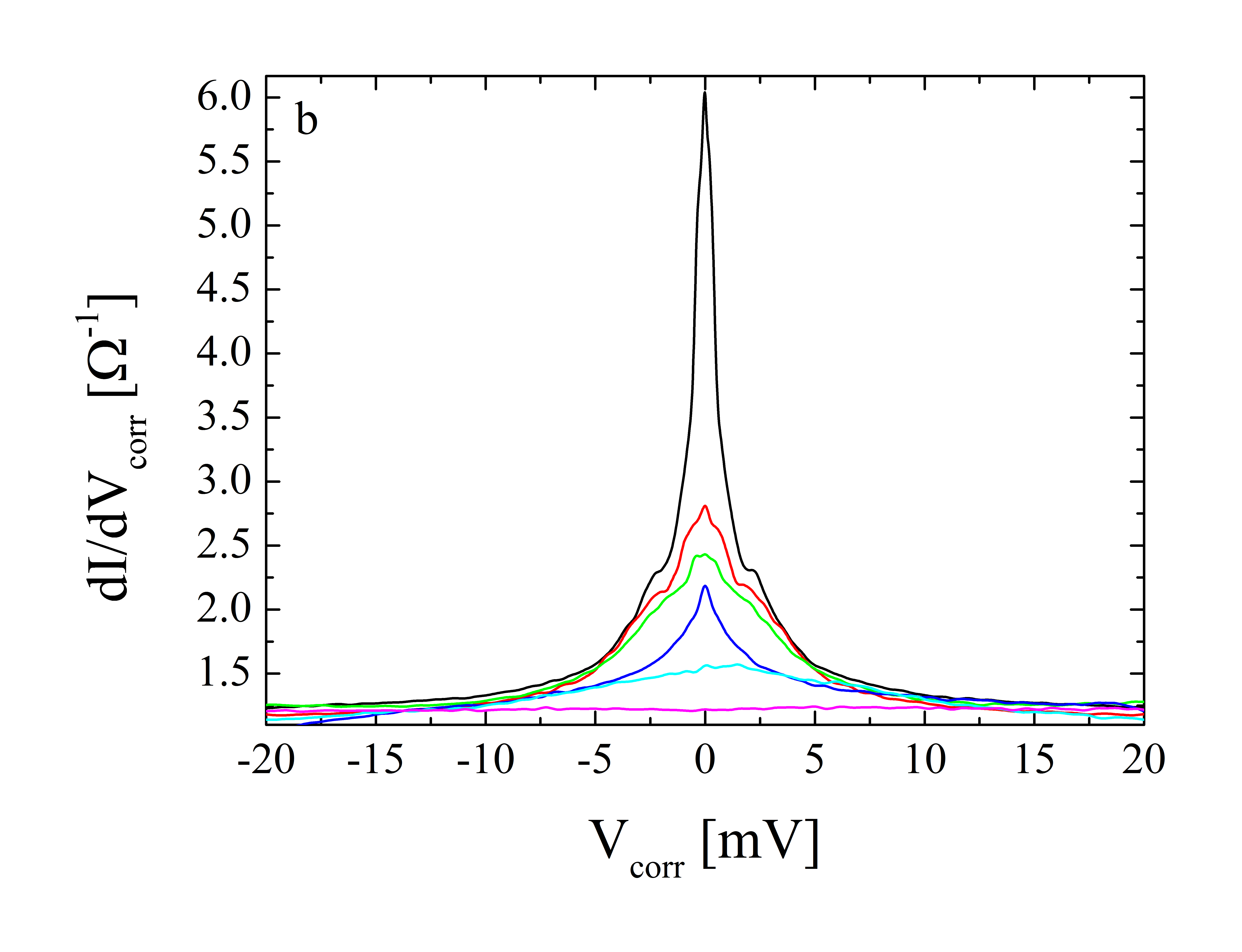}
\end{minipage}
\caption{\label{fig:G}Differential conductance of the same junction as figure~\ref{fig:R-I_uncorr} as measured (a) and corrected with equations~(\ref{eq:Vcorr}) and (\ref{eq:Rcorr}) as described in the text (b) for 4.2\,K, 8.0\,K, 12.2\,K, 16.6\,K, 20.9\,K and 23.0\,K.}
\end{figure}

The analysis of a superconductor order parameter is performed in the presentation of differential conductance versus voltage as shown in figure~\ref{fig:G}a. One can see, that the conductance for high voltages at $T$\,=\,4.2\,K goes to 1\,$\Omega^{-1}$ corresponding to the resistance of 1\,$\Omega$ in figure~\ref{fig:R-I_uncorr}. For higher temperatures one can see the influence of the electrode's resistance by a downshift of the high voltage value lowering as well as broadening of the central conductance peak. While the lowering of the peak and the downshift are connected to a loss of conductance the broadening is due to additional voltage at the electrodes. By using equation (\ref{eq:Rcorr}) to correct the conductance and (\ref{eq:Vcorr}) to correct the voltage one gets the actual values of the junction at each temperature without the influence of the electrode. The obtained spectra are shown in figure~\ref{fig:G}b. By comparing both spectra before and after the correction one can see, that there is of course no difference at $T$\,=\,4.2\,K but a noticeable difference for higher temperatures. Now all conductance curves merge at the same value of $G\approx1.0\,\Omega^{-1}$ at large voltage and the central peak broadens no more but stays nearly constant in width by changing temperature from 4.2\,K to 8.0\,K. There is still a loss in its height but this is natural due to the transformation from a SNS' junction to a SNN' junction by exceeding the critical temperature of lead.

\ack
This work was funded by the DFG within priority program 1458 (project no. SE 664/15-2), by the EC within project IRON-SEA (project no FP7-283141) and by the Landesgraduiertenf\"{o}rderung Th\"{u}ringen. The authors thank Renato Gonnelli and Dario Daghero from Politecnico di Torino for fruitful discussion.

\section*{References}
\bibliographystyle{iopart-num}
\bibliography{references}

\providecommand{\newblock}{}
\begin{thebibliography}{10}
\expandafter\ifx\csname url\endcsname\relax
  \def\url#1{{\tt #1}}\fi
\expandafter\ifx\csname urlprefix\endcsname\relax\def\urlprefix{URL }\fi
\providecommand{\eprint}[2][]{\url{#2}}

\bibitem{Blonder1982}
Blonder G~E, Tinkham M and Klapwijk T~M 1982 {\em Phys. Rev. B\/} {\bf 25}
  4515--4532

\bibitem{Dynes1978}
Dynes R~C, Narayanamurti V and Garno J~P 1978 {\em Phys. Rev. Lett.\/} {\bf 41}
  1509--1512

\bibitem{Plecenik1994}
Plecenik A, Grajcar M, Benacka S, Seidel P and Pfuch A 1994 {\em Phys. Rev.
  B\/} {\bf 49} 10016--10019

\bibitem{Kashiwaya1996}
Kashiwaya S, Tanaka Y, Koyanagi M and Kajimura K 1996 {\em Phys. Rev. B\/} {\bf
  53} 2667--2676

\bibitem{Mazin1999}
Mazin I~I 1999 {\em Phys. Rev. Lett.\/} {\bf 83} 1427--1430

\bibitem{Brinkman2002}
Brinkman A, Golubov A~A, Rogalla H, Dolgov O~V, Kortus J, Kong Y, Jepsen O and
  Andersen O~K 2002 {\em Phys. Rev. B\/} {\bf 65} 180517

\bibitem{Woods2004}
Woods G, Soulen R, Mazin I, Nadgorny B, Osofsky M, Sanders J, Srikanth H,
  Egelhoff W and Datla R {2004} {\em Phys. Rev. B\/} {\bf {70}} 054416

\bibitem{Chen2010}
Chen T~Y, Huang S~X and Chien C~L {2010} {\em Phys. Rev. B\/} {\bf {81}} 214444

\bibitem{Sheet2004}
Sheet G, Mukhopadhyay S and Raychaudhuri P 2004 {\em Phys. Rev. B\/} {\bf 69}
  134507

\bibitem{Baltz2009}
Baltz V, Naylor A~D, Seemann K~M, Elder W, Sheen S, Westerholt K, Zabel H,
  Burnell G, Marrows C~H and Hickey B~J 2009 {\em J. Phys.: Condens. Mat.\/}
  {\bf 21} 095701

\bibitem{Doering2012b}
D\"{o}ring S, Schmidt S, Schmidl F, Tympel V, Haindl S, Kurth F, Iida K,
  M\"{o}nch I, Holzapfel B and Seidel P {2012} {\em Physica C\/} {\bf {478}}
  15--18

\bibitem{Schmidt2012}
Schmidt S, D\"{o}ring S, Tympel V, Schmidl F, Haindl S, Iida K, Holzapfel B and
  Seidel P 2012 {\em Phys. Proc.\/} {\bf 36} 82--87

\bibitem{Iida2010}
Iida K, Haindl S, Thersleff T, Hanisch J, Kurth F, Kidszun M, Huhne R, Monch I,
  Schultz L, Holzapfel B and Heller R 2010 {\em Appl. Phys. Lett.\/} {\bf 97}
  172507

\end{thebibliography}

\end{document}